\documentclass[final,5p,times,twocolumn, preprint]{elsarticle}
\biboptions{sort&compress}
\usepackage{amssymb}
\usepackage{latexsym}
\usepackage{textcomp}
\usepackage{amsthm}
\usepackage{amsmath}
\usepackage{subfigure}
\usepackage{hyperref}
\usepackage{xcolor}
\usepackage{amssymb}

\journal{Radiation Measurements}

\begin{document}

\begin{frontmatter}



\title{Role of electron and hole centers in energy transfer in BaBrI crystals}

\author[label1]{Roman Shendrik}\ead{r.shendrik@gmail.com}
\author[label1]{Alexandra Myasnikova}
\author[label1]{Alexey Rupasov}
\author[label1]{Alexey Shalaev}

\address[label1]{664033, Vinogradov Institute of  Geochemistry SB RAS, 1a Favorskii street, Irkutsk, Russia}

\begin{abstract}

In this paper we study a role  of  F-centers,  hole  centers  and  excitons  in  energy  transfer in Eu-doped BaBrI crystals. Optical absorption spectra, thermally stimulated (TSL) and photostimulated (PSL) luminescence in wide temperature range 7-300 K are studied in undoped and doped with different concentrations of Eu ions BaBrI crystals. Based on experimental and calculated results two possible energy tranfer processes from host to Eu$^{2+}$ ions are established.
\end{abstract}


\end{frontmatter}


\section{Introduction}
Mixed alkaline earth halides BaBr$Y$ (where Y=I, Cl) are recently developed scintillation materials having good potential. The light output for BaBrI:Eu$^{2+}$ crystals was estimated as 90,000 photons/MeV \cite{bourret2012}, whereas in BaBrCl it was about 52,000 photons/MeV, respectively \cite{yan2016czochralski}. Additionally, they are less hygroscopic than  LaBr$_3$ and may have a broad implementation \cite{gundiah2011structure}. Theoretically these crystals could reach a higher luminosity, therefore further optimization toward improvements in the scintillation properties of this material should be possible. Thereby, it is necessary to study excitons \cite{shendrik2017optical, Shalaev201884}, electron and hole centers and their role in energy transfer to emission centers. 

In this paper we discuss the role of excitons, electron and hole centers in energy transfer process in BaBrI crystals. The results obtained by optical absorption, photostimulated and thermally stimulated luminescence are presented and discussed. The obtained results demonstrate possible energy transfer processes from host to Eu$^{2+}$ ions. Explanations for the role of electron and hole traps are presented as well.
\section{Methodology}

Eu-doped crystals and undoped crystals were grown by Bridgman technique reported in \cite{shendrik2017optical}. To study transformation of one type of F-centers into another, depending on the preparation conditions of the BaBrI we applied a Czochralski crystal growth technique. A conventional growth setup with a graphite thermal screens and resistivity heater was used. The crystals were grown in an inert (argon) atmosphere in glassy carbon crucible. For all growth experiments we used the seeds cut along from the crystal grown by Stockbarger method. The crystal pulling rate was 0.5 mm/hour, and rotation rate was 5 rpm.

Despite the fact that the quasi-binary phase diagram of BaI$_2$--BaBr$_2$ shows region where solid solution for either BaI$_2$ or BaBr$_2$ in BaBrI is not formed, an imbalance induced on purpose in the bromine/iodine ratio can alter the shape of the absoprtion spectrum from F(Br) to F(I)-centers. As shown below using imbalance between BaBr$_2$ and BaI$_2$ in raw the undoped crystal containing mostly F(I) centers can be grown. Further in this paper this sample is called non-stoichiometric crystal.

Optical absorption spectra were measured with a Perkin Elmer Lambda 950 spectrophotometer. The crystals were irradiated either by x-rays from a Pd tube operating at 40 kV and 40 mA regime for not more than one hour or 185 nm light from a low-pressure Hg-lamp. Absorption spectra of x-ray and photo colored crystals contain the same bands. Thermally stimulated (TSL) and photostimulated (PSL) luminescence spectra were measured in vacuum cold-finger cryostat. Before measurements samples were irradiated under 185 nm light from low-pressure mercury lamp at 77 K (for PSL) and at 7 K (for TSL) temperatures. PSL was registered with SDL-1 grating monochromator equipped with Hamamatsu photomodule H10721-04. PSL was excited using 150 W halogen lamp and MDR2 grating monochromator. TSL glow curves were measured in linear heating regime with rate 1 K/min using cryocooler Janis Research CCS-100/204N.

The calculations have been performed in embedded cluster approach implemented in Gaussian 03 computer code \cite{g03}. The QM cluster with vacancy and nearest neighbors (four or five barium ions) was surrounded by point charges. We used SDD basis with pseudopotential on barium ions and did not use any basis for vacancies, because it is shown that for the correct description of the F-center there are enough d-functions of the surrounding cations \cite{Mysovsky2011}. For optical absorption spectra calculations the time depended DFT (TD DFT) method was used. 

\section{Results and discussion}

\subsection{F-centers}

\begin{figure}[t!]
\centering
\includegraphics[width=0.5\textwidth]{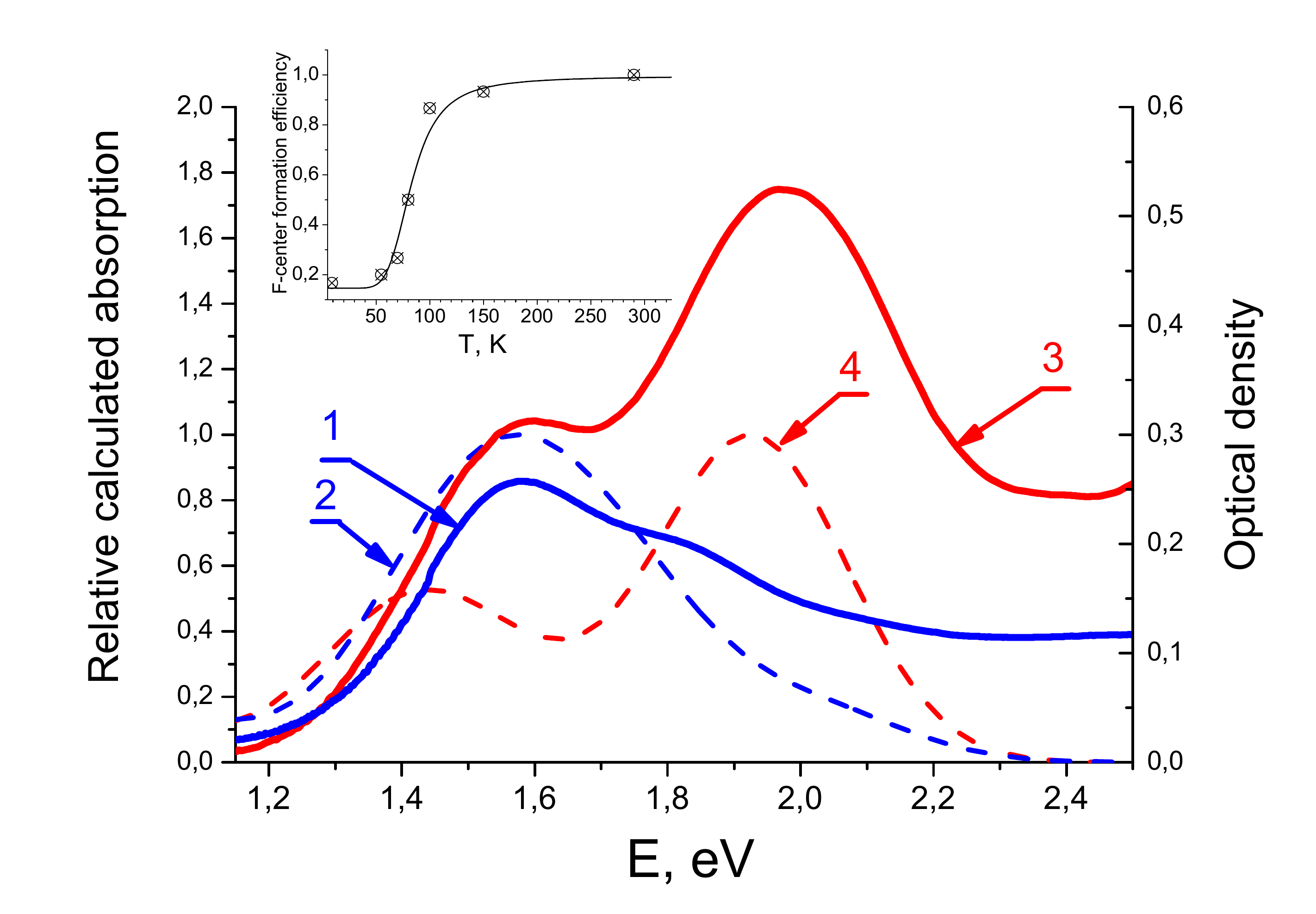}
\caption{Optical absorption spectra of nominally undoped BaBrI crystals irradiated at room temperature. Curves 1 and 3 are experimental spectra of  irradiated non-stoichiometric and stoichiometric crystals. Curves 2 and 4 are calculated spectra of F(I) and F(Br) centers. In the inset efficiency of F-centers creation is shown}
\label{absorption}
\end{figure}

After irradiation of undoped stoichiometric samples at room temperature bands at about 2.05 and 1.55 eV appear in optical absorption spectrum (Fig.~\ref{absorption}, curve 3). Intensity of band peaked at about 2.05 eV is higher than one of 1.55 eV band. In crystals grown using imbalance composition of BaI$_2$ and BaBr$_2$ absorption spectrum is changed. Higher energy band shifts to lower energy and its intensity becomes lower whereas intensity of low energy band is increased (Fig.~\ref{absorption}, curve 1). In irradiated Eu-doped samples these bands are not observed.

As shown in the inset of Fig.~\ref{absorption}, efficiency of F-centers creation decreases during cooling at temperatures about 100 K where intensity of STE band emission grows up \cite{shendrik2017optical, Shalaev201884}. This anticorrelation behaviour is similar to observed in alkali halides crystals \cite{PhysRevB.34.4230}.

In related crystals of BaFBr, BaBr$_2$ the absorption bands in irradiated crystals at about 2--2.5 eV are attributed to F(Br)-center \cite{radzhabov1995optical, houler, iwabuchi1991photostimulated}. Therefore, the observed bands could be assigned to F-centers. There exist one family of Ba$^{2+}$ ions and two families of anion ions. Since two types of halogen ions exist in these crystals, two different types of F centers may be expected. These are related to iodine F(I) and bromine F(Br) vacancies trapped an electron.

Ab initio calculation of F(I) and F(Br) centers were performed. The results are given in Fig.\ref{absorption}, curves 2 and 4, respectively. In calculated spectra several bands for each type of F-centers are observed. This is due to the fact that the ground state of F centers has approximately an $s$-like wavefunction. Whereas, the triple degeneracy of the excited $p$-states is partly lifted by the crystal field. The excited states split into three states due to monoclinic group C$_S$. In absorption spectrum of F(Br) centers the most intensive band is located at about 2~eV. For F(I) centers lower energy band at about 1.55~eV becomes the most intensive.

The calculated and experimental spectra agree well. In stoichiometric crystals two types of centers: F(Br) and F(I) are found. In non-stoichiometric samples grown by Czochralski method we observed an excess F(I) centers and optical absorption peak shifts to low energy region. 

\subsection{Photostimulated luminescence}

\begin{figure}[t!]
\centering
\includegraphics[width=0.5\textwidth]{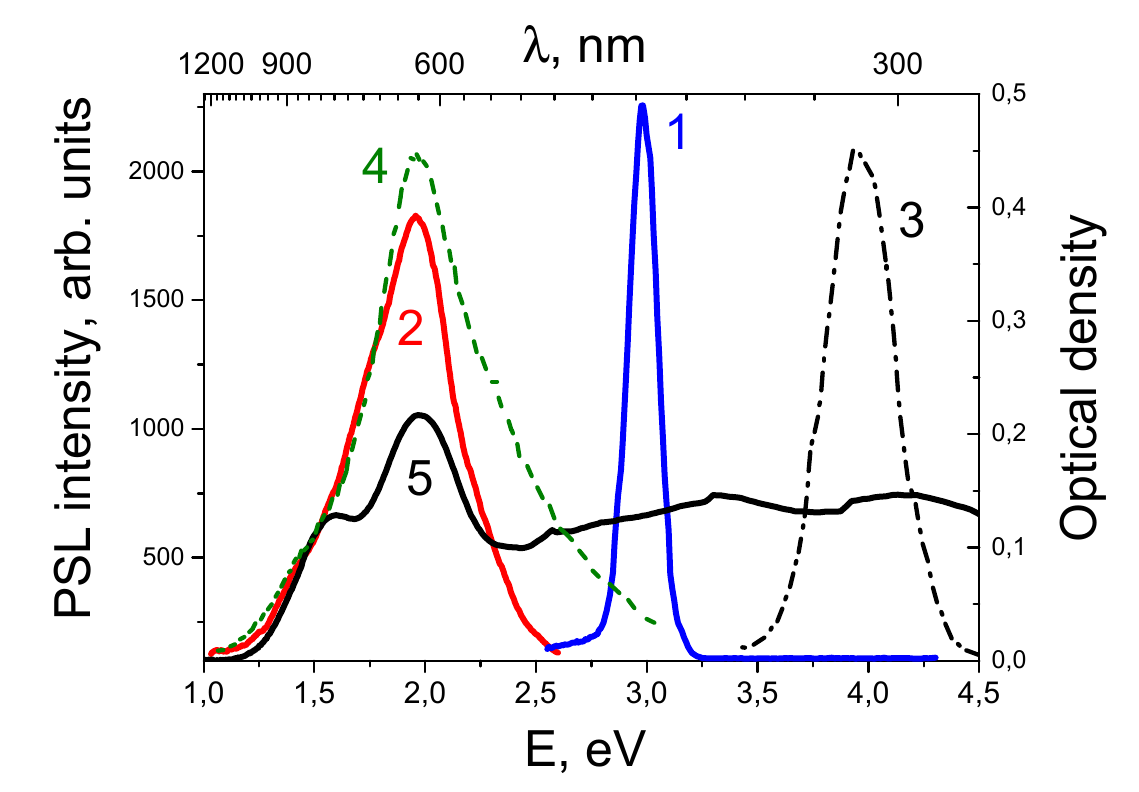}
\caption{Photostimulated luminescence (curve 1), excitation of PSL (curve 2) of BaBrI-0.05 mol.\% Eu crystal; and photostimulated luminescence (curve 3), excitation of PSL (curve 4) and optical absorption (curve 5) spectra of undoped BaBrI. All samples were irradiated at 77 K.}
\label{psl}
\end{figure}
When BaBrI doped with Eu sample is exposed to x-ray or 185~ nm irradiation at 77 K photostimulated luminescence is observed. Spectrum of photostimulated luminescence under 2 eV light irradiation is shown in Fig.~\ref{psl}, curve 1. The band peaked at about 3 eV in PSL spectrum is attributed to 5d-4f emission of Eu$^{2+}$ ions \cite{shendrik2017optical}. The excitation spectrum of this PSL is given in Fig.~\ref{psl}, curve 2. This spectrum shows peak at about 2 eV in the region where absorption of F(Br) centers is observed, see Fig.\ref{absorption} and Fig.~\ref{psl}, curve 5. Similar low temperature PSL has been observed in BaBrCl-Eu crystals \cite{SHALAPSKA2018497}. At temperatures higher than 120 K intensity of PSL is dramatically decreased.

Photostimulated luminescence is also observed in irradiated undoped crystals. The photostimulated luminescence spectrum is depicted in curve 3 of Fig. \ref{psl}. PSL peak at about 3.9~eV is attributed to STE exciton emission \cite{shendrik2017optical}. Photostimulation spectrum of exciton related PSL shows wide peak at about 2 eV (curve 4, Fig. \ref{psl}) similar to Eu-doped samples.
  
\subsection{Thermally stimulated luminescence}

\begin{figure}[t!]
\centering
\includegraphics[width=0.5\textwidth]{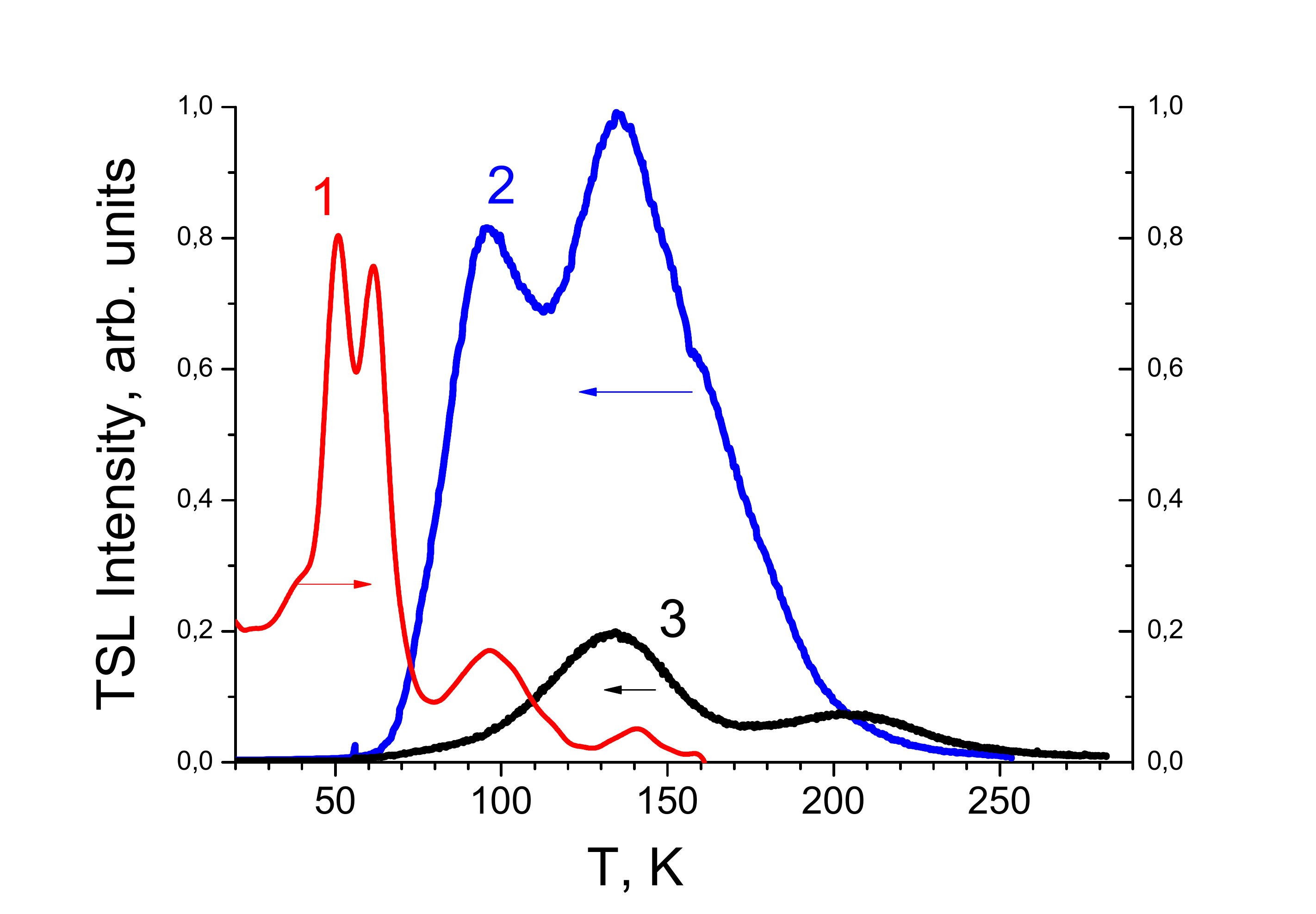}
\caption{TSL glow curves monitored in STE band (320 nm) (red curve 1) in undoped BaBrI and  5d-4f Eu$^{2+}$ band (420 nm) in BaBrI doped with 0.05 mol.\% Eu (blue curve 2) and 1 mol. \% Eu (black curve 3)}
\label{tsl}
\end{figure}

In addition to photostimulated luminescence, thermally stimulated luminescence properties of BaBrI and BaBrI doped with 0.05 and 1 mol.\% were investigated to clarify the recombination mechanism. In figure\ref{tsl}, curve 1 the glow curve for the exciton emission around 320 nm for x-ray irradiated at 7 K undoped BaBrI is depicted. Strong peaks around 50 and 60 K and two weaker peaks at 95 K and 140 K are observed. The symmetrical shape of the peaks indicates the presence of second or general order kinetics. For 50 and 60 K peaks thermal trap depth was estimated using second-order kinetic peak shape. The values are 0.10 and 0.13~eV respectively. Higher energy peaks are more complex and simple deconvolution is not possible.

The glow curves for Eu-doped samples are different to undoped sample. In figure\ref{tsl}, curve 2 the glow curves for the Eu$^{2+}$ emission around 410 nm for x-ray irradiated at 7 K BaBrI-Eu crystals are shown. Intense wide peaks at 95 and 140 K are found. In 1 mol.\% Eu$^{2+}$ doped sample intensity of TSL is weaker and two wide peaks at about 140 K and 200 K are observed (Fig.\ref{tsl}, curve 3). After irradiation in all samples hyperbolic afterglow is observed.

For all crystals the peaks in the glow curves correspond to beginning of thermally stimulated motion of hole centers, because F-centers are stable up to 600~K. Low temperature peaks at 50 and 60 K could be attributed to simple hole centers of I$_2^{-}$ or Br$_2^{-}$. Higher temperature peaks could be related to complex centers called in alkali halides V$_2$, V$_3$ centers \cite{Lushchik}. 

\subsection{Energy transfer to Eu$^{2+}$ ions}

\begin{figure}[t!]
\centering
\includegraphics[width=0.5\textwidth]{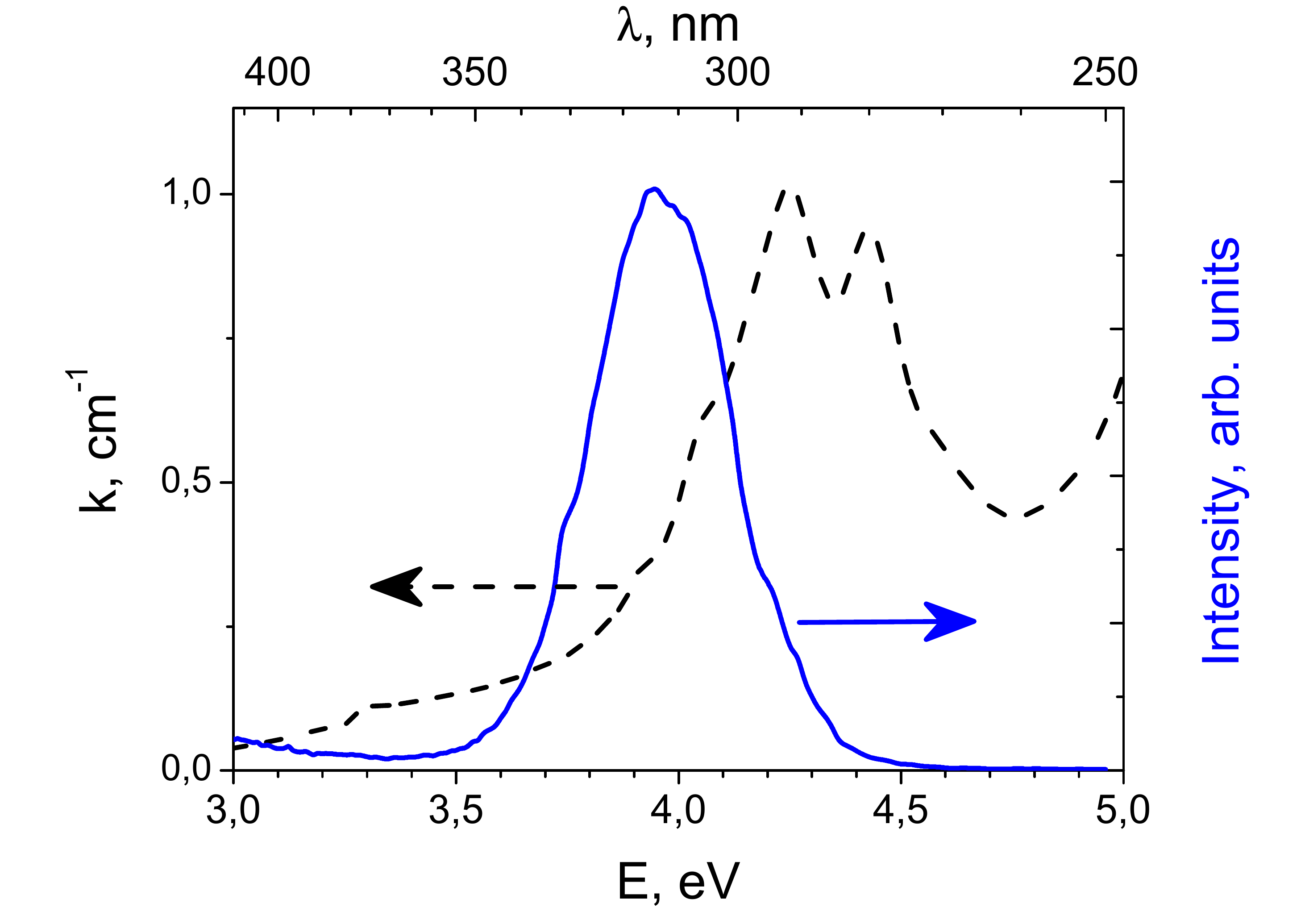}
\caption{Overlapping optical absorption of BaBrI-0.05 mol.\% Eu$^{2+}$ ions (dotted curve) and STE emission (solid curve) spectra}
\label{overlap}
\end{figure}

Excitation of Eu$^{2+}$ emission in the region of exciton peak \cite{Shalaev201884} can indicate the possibility for energy transfer from exciton to Eu$^{2+}$ ions as well as in LaBr$_3$:Ce, where resonance exciton energy transfer is dominant \cite{PhysRevB.97.144303}. Absorption of 4f-5d transition of Eu$^{2+}$ ions and exciton emission spectra overlap in BaBrI crystals (Fig.\ref{overlap}). Therefore dipole-dipole energy transfer from exciton to Eu$^{2+}$ ion is possible. The decrease in the exciton luminescence intensity accompanying the increase of Eu$^{2+}$ concentration is observed. At the level of 0.1 mol.\% of Eu$^{2+}$ the exciton emission is almost completely suppressed. Considering the uniform Eu$^{2+}$ ions distribution in the host lattice, the radius of resonant energy transfer from exciton to Eu$^{2+}$ ions, at which exciton emission is suppressed, can be estimated as half the distance between Eu$^{2+}$ ions. A half Eu-Eu distance at the level of 0.1 mol.\% is about 29~\AA.

This radius can be also estimated from overlapping emission and absorption spectra integral. The radius $R$ of dipole-dipole energy transfer defined as the distance at which the probability of donor (exciton in our case) radiative transitions is equal to the probability of transfer to acceptor (Eu$^{2+}$ ion), is given by:

\begin{equation}
R=\frac{B}{n^4 N_A}\int\limits_0^\infty\frac{f_D(E)\mu_A(E)}{E^4}dE
\end{equation}

Here $n$ is the refractive index of crystal, the subscript $A$ denotes the acceptor center and $D$ denotes the donor center, $N_A$ is concentration of acceptor centers (in cm$^{-3}$ ) and $\mu_{A}(E)$ is their absorption coefficient (in cm$^{-1}$), $f_{D}(E)$ is the emission spectrum of the donor centers after $\int\limits_0^\infty D(E)dE=1$ normalization. Constant $B=3h^4c^4/4\pi$ is equal to $3.7\cdot10^{-20}$ (eV$^4\cdot$cm$^4$). From the experimental data we estimate the radius $R$ as 27.6 \AA.

Both values obtained for the distance of energy transfer from the host to the Eu$^{2+}$ ion are in good agreement. Therefore, efficient energy transfer from host to Eu$^{2+}$ ion takes place. It explains the growing light output with concentration of Eu$^{2+}$ ions found in \cite{bourret2010}, because distance between Eu-Eu pairs is decreased with increasing of Eu$^{2+}$ ions concentration.
 
BaBrI optically excited decay curves were fitted with only one single exponential decay component \cite{shendrik2017optical, bizarri2011scintillation}. However, the decay curve under x-ray excitation was well approximated with three exponential decay components \cite{bizarri2011scintillation}. This fact, together with the presence of TSL, indicates the presence of delay energy transfer processes in the crystals \cite{shendrik2012energy}, when hole or electron traps take place in energy transfer from host to emission center.

In irradiated at 77 K samples PSL is observed. In stimulation spectrum of undoped crystal the band attributed to F(Br) centers appears. Taking into account the fact that only hole centers are unstable below room temperature we can propose following model. After irradiation free anion exciton decays to F-H pairs similarly to alkali halide crystals where anticorrelation between F-center production and intensity of exciton emission takes place \cite{song2013self, Lushchik}. Here F is F-center and H is H-center, which can be regarded as an X$_2^{-}$ molecule occupying an X$^{-}$ anion site. The anions concerning here can be iodine or bromine. H-centers become movable at relatively low temperatures about 40 K and they can recombine in thermally assisted process with F-centers causing TSL. Also they can form more complex (BrI)$^-$ hole centers. They are relatively stable at 77 K. Stimulation by light in absorption band of F(Br) centers causes that an electron is liberated from an F-center through conduction band and recombine with a hole center with exciton related luminescence. Afterglow is explained by tunneling recombination between an electron from F-center and hole.

In Eu-doped samples the delayed energy transfer process is more involved. From the one hand, Eu ions should be a hole trap relying on PSL data, where an electron is liberated from F-centers and radiatively recombine with hole. The simple assumption is that Eu$^{2+}$ ion captures a hole and becomes Eu$^{3+}$ ion. However similarly to BaFBr \cite{Seggern} and CsBr \cite{HACKENSCHMIED2002163} no Eu$^{3+}$ traces are found in irradiated crystals. Furthermore, TSL glow curves demonstrate that Eu$^{2+}$ should be an electron trap, because glow peaks correspond to hole traps which in thermally assisted process recombine with an electron on Eu ion. So, we conclude that as the case for BaFBr-Eu and CsBr-Eu the hole trap contains Eu$^{2+}$ ion with a hole nearby. 

As the case BaFBr crystals F-centers and H-centers formation can be spatially correlated and appear near Eu$^{2+}$ ion \cite{Koschnick} with further stabilization a hole centers by Eu$^{2+}$ ion. When an electron released from an F-center recombines with a hole, Eu$^{2+}$ emission excited. From the other hand, a hole became thermally unstable can recombine with an electron in F-center and Eu$^{2+}$ emission is also excited. This process is more effective in crystals doped with high Eu$^{2+}$ concentrations due to smaller distances between Eu ions. Therefore, long time components make lower contribution to luminescence decay. 

\section{Conclusion}

STE/Eu$^{2+}$ emission/absorption overlap allows to determine dipole-dipole transfer distance. The distance is sufficient to resonant energy transfer. This concludes that the fast energy transport from host to activator responsible for the scintillation of BaBrI-Eu proceeds by STE creation and resonance dipole-dipole transfer. At the same time, delayed energy transfer with participation of electron and hole centers formed by exciton decay takes place. In this process spatially correlated pair of F- and hole-center is produced, and a hole recombines with F-center in thermally assisted process leading to resonance transfer to Eu ion. Contribution of delayed energy transfer process decreases with increase of concentration of Eu$^{2+}$ ions due to diminution of distance between the dopant ions.



\section*{Acknowledgments}

Crystal growth of Eu-doped samples were supported by Russian Academy of Science project 0350-2016-0024. All spectroscopic data, calculations and crystal growth of undoped samples were supported by grant of Russian Science Foundation RSF 17-72-10084. In this work author used the equipment of the Isotopic and Geochemistry
Research Centre for Collective Use, Vinogradov Institute of Geochemistry, Russian Academy of Sciences.

\bibliography{photochromic}
\end{document}